Abstract: This paper presents an evaluation of the WiFi exposure levels inside the university in the 2.4 GHz frequency band. The selected environment is the typical scenario where WiFi exposure concerns have increased in the last years, since a Wireless Local Area Network is deployed close to the users. Measurements of 1 h and 24 h duration were performed to assess the temporal and spatial variability of the signal. Two instruments were employed, a spectrum analyzer appropriate configured for recording accurate and realistic samples and an exposimeter. A detailed description of the equipment, the measurement procedure and data analysis is provided in order to allow the reproducibility of these types of measurements. Finally, a comparison of the WiFi levels obtained by other authors is presented, concluding that all these methods are useful for determining WiFi exposure distribution, but if more accurate results are required, professional equipment appropriately configured should be used.

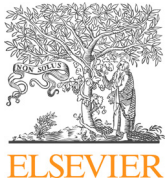

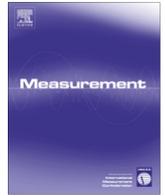

# Measurements and analysis of temporal and spatial variability of WiFi exposure levels in the 2.4 GHz frequency band

Marta Fernández *, David Guerra, Unai Gil, Iñigo Trigo, Ivan Peña, Amaia Arrinda

*Communications Engineering Department, University of the Basque Country (UPV/EHU), Alda. Urquijo s/n, 48013 Bilbao, Spain*

ABSTRACT

This paper presents an evaluation of the WiFi exposure levels inside the university in the 2.4 GHz frequency band. The selected environment is the typical scenario where WiFi exposure concerns have increased in the last years, since a Wireless Local Area Network is deployed close to the users. Measurements of 1 h and 24 h duration were performed to assess the temporal and spatial variability of the signal. Two instruments were employed, a spectrum analyzer appropriate configured for recording accurate and realistic samples and an exposimeter. A detailed description of the equipment, the measurement procedure and data analysis is provided in order to allow the reproducibility of these types of measurements. Finally, a comparison of the WiFi levels obtained by other authors is presented, concluding that all these methods are useful for determining WiFi exposure distribution, but if more accurate results are required, professional equipment appropriately configured should be used.

© 2019 Elsevier Ltd. All rights reserved.

## 1. Introduction

The assessment of radiofrequency (RF) fields is a great concern in our society and the data collection by means of robust methods is a need in order to quantify the exposure levels and its variability. The good knowledge of electromagnetic field exposure levels is essential for two main reasons: give response to public concern, and ensure people's protection against these emissions but without reducing the technological benefits because of overly restrictive deployment policies.

Several measurement campaigns were performed with the aim of evaluating electromagnetic field levels in different environments. In some of them, data were recorded from a wide range of frequencies without distinguishing between services [1–3]. Other authors identified emissions coming from different frequency bands, for example using personal exposure meters, which have the advantage of being portable and easy to use [4,5], or by means of more professional equipment such as spectrum analyzers [6–10]. When using exposure meters, many times different groups of people were volunteered to participate in the measurement campaign carrying the device with them in order to evaluate the personal exposure of the participants and, in some cases, they even filled questionnaires and diaries about the activities they realized during the duration of the study. Some works for characterizing human exposure to RF fields included several of the methods and equipment above described, such as in [11,12]. In addition, computer tools were also used in order to estimate electromagnetic fields from different sources [13]. Some authors focused on evaluating the exposure due to a specific service, such as mobile communication networks [14–18] or Wireless Local Area Networks (WLAN), which use WiFi technology [19,20].

Emissions specifically due to WiFi signals have caused concerns in the last years, especially because the transmitters are very close to people and WLAN are deployed in indoor environments, such as schools, homes or office buildings where people spend much time. Most of the devices using WiFi technology operate in the 2.4 GHz or 5 GHz WiFi bands. At 2.4 GHz transmission is set between 2.4 and 2.4835 GHz as allocated by regulatory bodies in China, United States, Europe and Japan, or also in the frequency range from 2.471 to 2.497 GHz in the case of Japan. These frequencies correspond to a part of the bands assigned for industrial, science and medical (ISM) applications, while at 5 GHz devices can operate in the ISM band (5.725–5.850) or at different frequencies (5.150–5.350 GHz). The original IEEE 802.11 standard for wireless local area networks was published in 1999 and reaffirmed in 2003. Later, some revisions have been published, the last one in 2016 [21].

Due to the nature of WiFi emissions, which are transmitted in the form of pulses of short duration, professional equipment is required in order to obtain accurate values of WiFi exposure [22]. Several WiFi measurement campaigns were carried out using exposure meters [4,5] or spectrum analyzers configured for obtaining worst-case exposure levels [8]. These two methods are useful for checking compliance with regulations or for having an idea of the electric field distribution, but not appropriate if more accurate levels are required. In such a case, spectrum analyzers properly configured are necessary. Some authors introduced weighting factors when employing spectrum analyzers in order to account for the variability of WiFi signals, but in these cases additional measurements have to be performed in order to calculate the weighting factors [23,24]. Recently, the authors of the present paper proposed a spectrum analyzer configuration for taking accurate WiFi signal samples using only frequency domain measurements [25]. In that work, measurements in controlled conditions were performed and a good agreement between the data traffic and the obtained values was observed.

In this paper, an evaluation of the exposure levels due to WiFi signals at 2.4 GHz inside the university is presented. One of the objectives is to provide accurate results of the temporal and spatial variability of the signal in the chosen environment, investigating the correlation between WiFi exposure variability and the different days of the week. The second objective is to show the methodology followed for obtaining and analyzing the results in order to allow the reproducibility of these types of measurements. Two different instruments were employed to assess WiFi exposure, a spectrum analyzer configured as described in [25] so as to take accurate signal samples and a personal exposure meter. Finally, a comparison is made between several WiFi measurement campaigns carried out in different studies.

## 2. WiFi exposure assessment

The access points of the university network were using the IEEE 802.11n standard, which allows transmissions in the 2.4 GHz or 5 GHz frequency bands. Power measurements were carried out in the 2.4 GHz band where the bandwidth of each WiFi channel is equal to 20 MHz. The maximum data rate supported by this standard for a 20 MHz channel is 288.9 Mbps [26]. In Europe, transmission is allowed in 13 different channels in this frequency band and the separation between them is 5 MHz, thus some channels overlap with each others. As specified in the standard [21], adjacent cells using different channels can operate simultaneously without interference if the distance between the center frequencies is at least 25 MHz.

### 2.1. Measurement scenario

Measurement campaigns were performed in a faculty of the University of the Basque Country (Spain), which is located in an urban environment surrounded by homes, offices and restaurants. This area is characterized by a high level of activity, next to the university there is a bus station and a soccer stadium. As shown in Fig. 1, the university includes several buildings, one called B next to the bus station, a narrower one (F), which passes over the highway and connects the parts B and D. Next to the latter there is another corridor called E that connects with the rest of the university: three parts named A, C and G, which form an inner courtyard. Measurements were carried out inside the university, in two labs, three classrooms and three corridors of different floors.

### 2.2. Types of measurements

Measurements to assess WiFi exposure were performed with the aim of evaluating the temporal and spatial variability of the signals. For this purpose, two types of tests were done: long-term measurements, in which samples were recorded continuously during 24 h at each point to study the variability of the signal throughout the day and 1-hour measurements at each point thus allowing the measurements at multiple points in the same WiFi activity conditions in order to characterize the spatial variability.

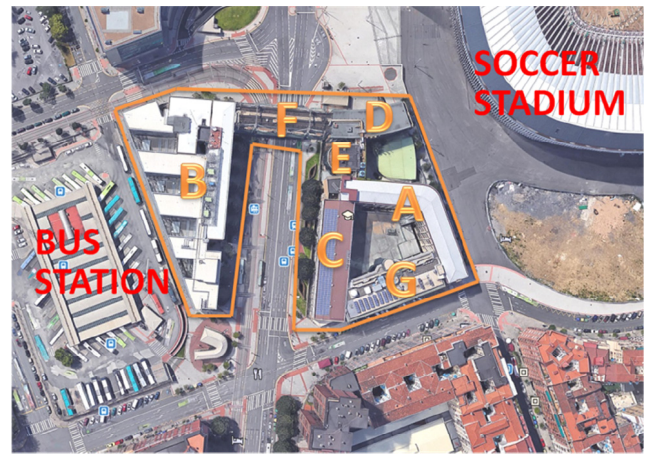

**Fig. 1.** Location of the university where measurements were carried out.

Long-term tests were performed in two labs on the 4th floor of building B. Measurements of 24 h duration were recorded at 5 different points in each lab: at the center and in the four corners of the room at 1.4 m (diagonally) from the corner, following the procedure used in [11,27]. The receiver instruments were placed at a height of 1.2 m above the floor; since in these labs students and researchers are usually sitting so this corresponds to the average height of their head. In relation to this, the Institute of Electrical and Electronics Engineers recommends measuring RF fields at a height not higher than 2 m above the floor [28]. Moreover, WiFi signals were evaluated two different days at each point in order to investigate the correlation between different days, so long-term recordings were done during a total of 10 days (240 h) in each lab.

1-hour measurements were carried out at six positions of each corridor and classroom. In the case of the classrooms, the receiver equipment was placed in the four corners, at a distance of 1.4 m from them to avoid walls influence and at two locations close to the center of the room. The height of the receiving antennas was 1.2 m above the floor, as students are usually sitting in these places. In the corridors, evaluation points were in the middle of the two walls of the corridor at a height of 1.7 m, because people are usually walking or standing in these locations. A distance of 3 m between two consecutive measurement points was chosen. Table 1 summarizes the different places where samples were recorded, indicating the building and the floor of each measurement location, the measurement duration (Meas. Time) and the number of positions selected at each location (N° of Pos.). Finally, the total recording time at each place is given. As in Lab 1 and Lab 2 samples were taken two different days at each point, a total of 10 measurements were performed in each lab. In the corridors and classrooms, one measurement per position was taken. 1-hour measurements were performed during the weekdays and the three classrooms were empty during the measurements, but beacon signals were received from the closest access points as well as data traffic signals due to people's devices on the corridor or in other classrooms. 24-hour measurements were taken during the 7 days of the week and part of the time people were in the labs while samples were recorded. The university is open on weekdays from 7:30 to 21:30 and on Saturday morning.

**Table 1**
Description of the measurements.

| Place | Building/floor | Meas. time | N° of Pos. | Total time |
|---|---|---|---|---|
| Lab 1 | B/4 | 24 h | 5 | 240 h |
| Lab 2 | B/4 | 24 h | 5 | 240 h |
| Corridor 1 | B/3 | 1 h | 6 | 6 h |
| Corridor 2 | A/2 | 1 h | 6 | 6 h |
| Corridor 3 | B/2 | 1 h | 6 | 6 h |
| Classroom 1 | C/3 | 1 h | 6 | 6 h |
| Classroom 2 | B/3 | 1 h | 6 | 6 h |
| Classroom 3 | G/1 | 1 h | 6 | 6 h |

*2.3. Equipment*

Two different types of instruments were used to measure human exposure to WiFi signals. The differences in the obtained results due to the measurement instrumentation were evaluated. As professional equipment, an EMI ESPI3 spectrum analyzer of Rohde & Schwarz that works in the frequency range from 9 kHz to 3 GHz was utilized [29], together with a tri-axial antenna system composed of three Yagi antennas (model ANYG1502) suitable for the 2.4 GHz WiFi band. The maximum antenna gain given by the manufacturer was equal to 15 dBi. We measured antenna gain at the center frequency of the WiFi band and it was equal to 12 dBi, so this is the value for the antenna gain used in this work. The antennas were placed in three mutually orthogonal directions and were connected using two combiners and an attenuator, so the power received by the three antennas experienced the same losses. The total losses due to the combiners and to the cables were then added manually.

Some concerns may arise regarding the distortion of the radiation pattern of the antennas when setting up the tri-axial system. So, in order to validate it, the electric field strength measured by this system was compared with the electric field strength assessed using a second method. The latter consisted in measuring three times with one of the Yagi antennas placed in the three orthogonal directions each time ($E_x$, $E_y$, $E_z$). Then, the field strength values were combined using (1)

$$E_{Total} = \sqrt{|E_x|^2 + |E_y|^2 + |E_z|^2}$$ (1)

This comparison was made at night to ensure the same situation during the four measurements (the one with the tri-axial system and the three measurements required when using the second method). So a total of four nights were required to compare these systems. One sample per second was recorded and every hour the median value of the power measured was calculated. Fig. 2 shows the difference in the median power strength obtained by means of the both methods. The median of the power level measured by means of the tri-axial antenna system was always higher. As seen, the greatest difference was 1.14 dB obtained at 6:00 am. These differences are not only due to the antenna measurement system, but also to different activity in the WLAN, since measurements were performed in different days.

A computer wired connected to the spectrum analyzer was used to configure the analyzer and to save the recorded data. The spectrum analyzer settings can be seen in Table 2. These settings were tested in controlled conditions by the authors in a previous work and the main advantage is that accurate and realistic WiFi levels are obtained performing measurements only in the frequency domain [25]. In that work, a procedure based on both time and frequency domain measurements was followed in order to obtain the measurement configuration. First, a set of reference samples of the radiation caused by a perfectly known WiFi signal was obtained. Then, these reference samples were compared with the levels registered for the same type of signal considering different values of the spectrum analyzer parameters in order to analyze their influence on the measurements. It was concluded that the SWT was the most influential parameter and the error associated with each one of the frequency domain configurations was quantified, showing that a SWT equal to 2.5 ms produces the lowest error in the measured level.

The center frequency and span were set to 2441.75 MHz and 83.5 MHz, respectively, in order to detect the signals of the different channels in the 2.4 GHz WiFi band. The root mean square (RMS) detector and the clear/write trace were selected to avoid over and underestimations of the WiFi emissions. One sample was recorded each second in a text file for further processing.

The other measurement equipment used in the measurement campaigns was an exposimeter or personal exposure meter, the EME Spy 200 [30]. It was also configured to measure in the 2.4 GHz WiFi band (2400–2483.5 MHz) and a value of the electric field strength was recorded every 4 s. This device has a tri-axial electric field probe valid for the frequency range from 80 MHz to 6 GHz and the detection limit in the frequency range of interest is 0.005 V/m. For values lower than this limit the EME Spy uses the naïve approach, which consists of replacing the measured level by the detection limit [31].

All the measurements were performed using the spectrum analyzer together with the Yagi antenna system and in the case of the 1-hour measurements, also the personal exposure meter was employed. For these last measurements, both instruments were recording samples at the same time, being the minimum distance between them equal to 40 cm in order to ensure that one instrument was not in the near field region of the other receiver. As the receiving Yagi antenna is physically larger than half a wavelength, the far field region starts at the distance $d_f$, which is dependent of the largest dimension of the antenna $D$ and the wavelength $\lambda$ [32]:

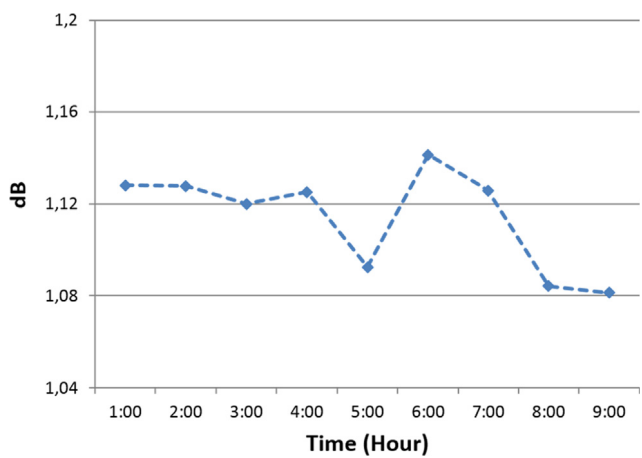

**Fig. 2.** Difference in the power levels measured using the two different antenna systems from 1:00 to 9:00 in the morning.

**Table 2**
Configuration of the spectrum analyzer.

| Parameter | Value |
|---|---|
| Centre Frequency | 2441.75 MHz |
| Span | 83.5 MHz |
| Detector | RMS |
| SWT | 2.5 ms |
| RBW | 1 MHz |
| VBW | 3 MHz |
| Trace Mode | Clear/Write |

$$d_f = \frac{2D^2}{\lambda} \qquad (2)$$

This distance was calculated using the lowest $\lambda$ in the 2.4 GHz WiFi band, which is 12.08 cm and the largest antenna dimension, which is equal to 15 cm corresponding to the Yagi antenna.

### 2.4. Data analysis and statistical discussion

The recorded data were saved in a text file for further processing. As above mentioned one sample per second was obtained when measuring with the spectrum analyzer and a sample every 4 s when using the personal exposure meter. Test of normality was applied to find out whether measurement samples were normally distributed or not, and results showed that data were not normally distributed. Regarding the 24-hour measurements, in order to determine exposure variations at different moments of the day, three different periods of time were distinguished: morning (6:00–14:00), afternoon and evening (14:00–22:00) and night (22:00–6:00). In addition, the correlation with the days of the week was statistically investigated.

Every hour several percentiles of the electric field strength were calculated as well as the minimum, maximum and mean values. Data statistics of each position inside a location (lab, classroom or corridor) were evaluated separately. Then, statistic results were calculated taking into account all the data acquired at one lab, classroom or corridor, without distinguishing the receiving position, so the mean WiFi exposure levels at each location were assessed, as well as the standard deviation. For example, the average 50th percentile during the morning period in a lab was obtained by calculating the average value of all the 50th percentiles acquired during the morning. These statistics were calculated in linear units, but for illustrating the results in graphs logarithmic units were chosen.

The 90th percentile was considered an appropriate statistic for representing the WiFi exposure variations because of the nature of these signals, which are transmitted in the form of bursts. The median value or 50th percentile has fewer variations along the day and between the different places. Percentiles higher than the 99th, which can have higher variations, are not representative of electromagnetic field exposure since they indicate singular occurrences of the signal. However, values such as the median or the maximum reached levels are significant statistics to evaluate exposure levels and to compare with regulations and standards. As explained in [33], median exposure values are more interesting for epidemiological studies, while maximum values are often more important for authorities and legislation.

## 3. Results and discussion

### 3.1. Long-term measurements

The results obtained using the spectrum analyzer and the triaxial antenna system in the two laboratories are presented below. Fig. 3a and Fig. 3b show the mean values of the 90th and 99th percentiles (P90, P99) of the electric field measured with the spectrum analyzer in Lab 1 and Lab 2, respectively. The results obtained in the four corners (Pos 1, Pos 2, Pos 3, Pos 4) and in the middle of each room (Pos 5) are represented for the different periods of the day. As shown, the recordings were repeated two days at each position.

Significant differences were observed due to the different positions inside the room. In Lab 1, the maximum difference between the P99 of the different positions was obtained during the morning periods and it was equal to 14.75 mV/m. For the P90, differences of up to 1.85 mV/m were found when placing the receiver at different points. In Lab 2, the P99 of the afternoon tests differed by 42.97 mV/m between positions 2 and 5, while the maximum variations of the P90 due to different positions were found during the morning and reached a value of 3.33 mVm. The variability at night was lower than during the day. When considering the P99, the levels of the night tests were higher than the tests carried out during the day in some of the measurements, this is because these percentiles indicate singular occurrences of the signal and they are not representative of WiFi exposure.

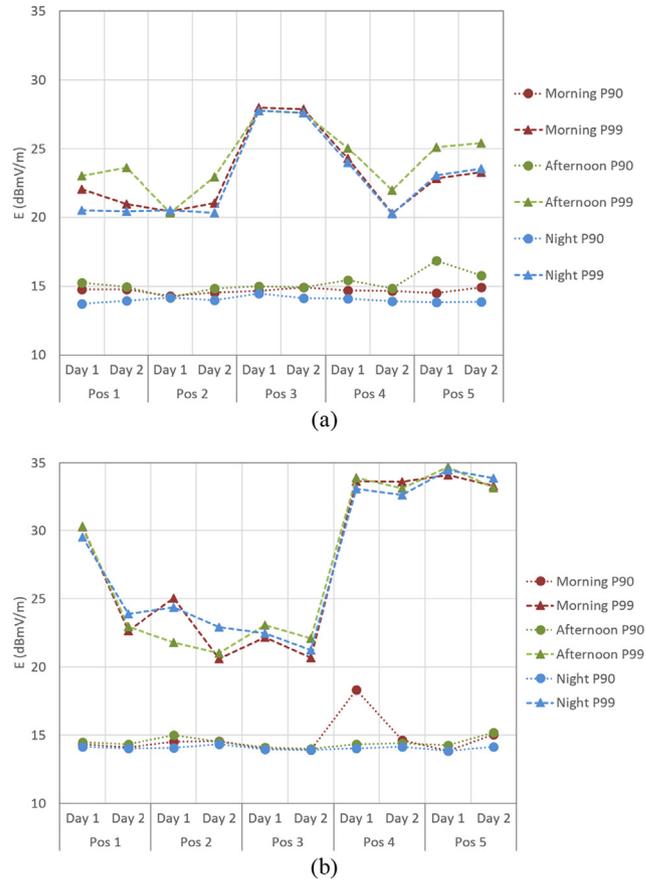

**Fig. 3.** 90th and 99th percentiles of the electric field levels measured at the different positions (a) in Lab 1, (b) in Lab 2.

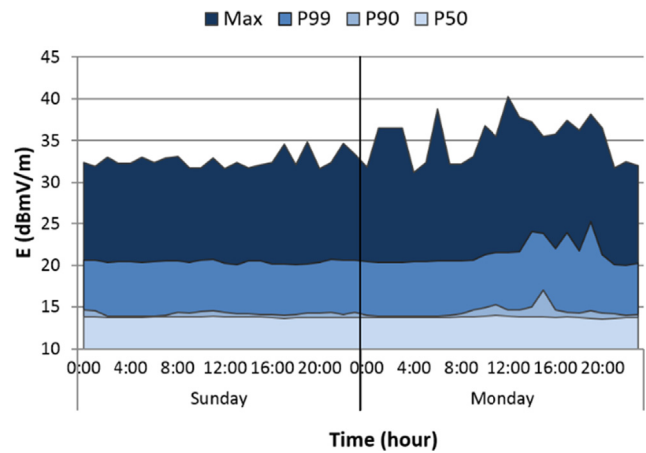

**Fig. 4.** Temporal evolution of the electric field strength measured during two consecutive days in the position 2 of Lab 1.

**Table 3**
Test of normality.

| Parameter | Value |
|---|---|
| W-stat | 0.413 |
| p-value | $1.644 \times 10^{-09}$ |
| Alpha | 0.05 |
| Normal | No |

Differences between the weekdays and the weekends were also found. An example of this is illustrated in Fig. 4, where the median (P50), P90, P99 and the maximum levels measured in the position 2 of Lab 1 are represented. The first 24-hour measurement at this point was taken on Sunday and the variability of the WiFi exposure was lower than in the measurement performed the following day. On Monday, the signal increased during the working hours at university.

The average WiFi exposure levels for each lab are summarized in Table 4. The mean electric field levels obtained in each lab during the different periods of the day are presented together with the standard deviation. The mean values of the different percentiles were calculated considering all the measurements taken during the same period of the day in the different positions of the lab. Finally, the minimum and maximum sample levels measured in the morning, afternoon and night are given. The mean value of the WiFi exposure was higher in Lab 2, being this value 5.25 mV/m in the mornings and afternoons. In both labs the electric field strength measured during the day was higher than the levels obtained at night. Regarding the average P90, in Lab 1 it took values between 5.02 mV/m (at night) and 5.77 mV/m (in the afternoon), while in Lab 2 it ranged between 5.05 mV/m (at night) and 5.51 mV/m (in the morning), the maximum standard deviation associated to these sets of calculations was 0.94. For the average P99, the field strength levels increased considerably, reaching a value of 29.21 mV/m in Lab 2. The maximum level of WiFi exposure in Lab 2 was equal to 407.81 mV/m, while in Lab 1 the maximum measured value was 172.26 mV/m, both of them recorded in the afternoon.

An example of the results obtained using Shapiro-Wilk test for testing the normality of the data measured in Lab 1 are presented in Table 3. A significant level (alpha) of 0.05 was established for statistical analysis and a Wilk-Shapiro test statistic (W-stat) of 0.413 was obtained. According to test of normality, data were not normally distributed as p-value was lower than 0.05.

Finally, Pearson correlation coefficient was performed to determine the correlation between WiFi exposure variability and the days of the week. The mean electric field levels obtained for the different hours of the day were used to investigate the linear relationship between the exposure variability in different days. Table 5 and Table 6 show the Pearson correlation coefficient in Lab 1 and Lab 2, respectively. The measurement position and the weekday is given in the tables and the first finding is that WiFi exposure variability in Lab 1 showed a more uniform behavior than in Lab 2. This is due to the activity in each lab, since the first lab is use for research purposes and people work there during the same hours on weekdays. However, the second lab is a teaching lab where students go to do laboratory practices and each day the lessons are at different hours. For this reason, in Lab 2 exposure of different days was less correlated, being the Pearson coefficient a negative value in some cases, since at the same hours WiFi levels increased one day and decreased on other day. The best correlation coefficient in this lab was equal to 0.829, while the largest correlation coefficient in Lab 1 was 0.927, reached between Position 2 on Monday and Position 1 on Thursday. The smallest correlation coefficient in Lab 1 was 0.008 obtained between Position 2 on Sunday and Position 5 on Monday, which makes sense because on Sunday the university is close, so WiFi signal variability is expected to be different than on weekdays.

**Table 4**
Average electric field strength values (mV/m) measured with the spectrum analyzer in each lab.

| | Lab 1 | | | Lab 2 | | |
|---|---|---|---|---|---|---|
| | Morning | Afternoon | Night | Morning | Afternoon | Night |
| Min | 4.58 | 4.49 | 4.57 | 4.59 | 4.61 | 4.62 |
| Mean (SD) | 5.17 (0.08) | 5.15 (0.09) | 5.03 (0.01) | 5.25 (0.17) | 5.25 (0.17) | 5.18 (0.10) |
| P50Av[a] (SD) | 4.89 (0.05) | 4.79 (0.05) | 4.82 (0.05) | 4.87 (0.05) | 4.84 (0.04) | 4.87 (0.05) |
| P90Av[a] (SD) | 5.41 (0.11) | 5.77 (0.47) | 5.02 (0.12) | 5.51 (0.94) | 5.29 (0.22) | 5.05 (0.08) |
| P99Av[a] (SD) | 15.05 (5.26) | 16.92 (4.41) | 14.59 (5.24) | 29.14 (16.72) | 29.21 (17.09) | 29.14 (15.99) |
| MaxAv[a] (SD) | 104.78 (36.10) | 117.06 (40.20) | 91.76 (40.05) | 185.95 (129.42) | 178.18 (124.94) | 177.90 (119.32) |
| Max | 154.82 | 172.26 | 149.88 | 388.77 | 407.81 | 380.16 |

SD: standard deviation.

[a] Mean values of the P50, P90, P99 and maximum levels of the electric field (mV/m) measured in the different positions of the lab.

**Table 5**
Pearson correlation coefficient for the mean electric field at the different hours of a day in Lab 1.

| | Pos 1 Thu | Pos 1 Tue | Pos 2 Sun | Pos 2 Mon | Pos 3 Sun | Pos 3 Sat | Pos 4 Tue | Pos 4 Fri | Pos 5 Mon | Pos 5 Wed |
|---|---|---|---|---|---|---|---|---|---|---|
| Pos 1 Thu | 1 | 0.855 | 0.324 | **0.927** | 0.656 | 0.717 | 0.849 | 0.768 | 0.571 | 0.895 |
| Pos 1 Tue | 0.855 | 1 | 0.454 | 0.801 | 0.669 | 0.577 | 0.880 | 0.850 | 0.540 | 0.874 |
| Pos 2 Sun | 0.324 | 0.454 | 1 | 0.286 | 0.546 | 0.273 | 0.466 | 0.471 | **0.008** | 0.301 |
| Pos 2 Mon | **0.927** | 0.801 | 0.286 | 1 | 0.643 | 0.646 | 0.858 | 0.769 | 0.698 | 0.910 |
| Pos 3 Sun | 0.656 | 0.669 | 0.546 | 0.643 | 1 | 0.774 | 0.608 | 0.771 | 0.529 | 0.635 |
| Pos 3 Sat | 0.717 | 0.577 | 0.273 | 0.646 | 0.774 | 1 | 0.509 | 0.508 | 0.454 | 0.553 |
| Pos 4 Tue | 0.849 | 0.880 | 0.466 | 0.858 | 0.608 | 0.509 | 1 | 0.811 | 0.524 | 0.847 |
| Pos 4 Fri | 0.768 | 0.850 | 0.471 | 0.769 | 0.771 | 0.508 | 0.811 | 1 | 0.705 | 0.856 |
| Pos 5 Mon | 0.571 | 0.540 | **0.008** | 0.698 | 0.529 | 0.454 | 0.524 | 0.705 | 1 | 0.668 |
| Pos 5 Wed | 0.895 | 0.874 | 0.301 | 0.910 | 0.635 | 0.553 | 0.847 | 0.856 | 0.668 | 1 |

Bold values correspond to the maximum and minimum values. This is, the best correlation coefficient and the smallest correlation coefficient.

**Table 6**
Pearson correlation coefficient for the mean electric field at the different hours of a day in Lab 2.

|  | Pos 1 Mon | Pos 1 Tue | Pos 2 Thu | Pos 2 Fri | Pos 3 Sun | Pos 3 Sat | Pos 4 Tue | Pos 4 Wed | Pos 5 Sat | Pos 5 Fri |
| --- | --- | --- | --- | --- | --- | --- | --- | --- | --- | --- |
| Pos 1 Mon | 1 | 0.590 | 0.012 | −0.009 | 0.408 | −0.470 | 0.573 | 0.701 | 0.436 | **0.829** |
| Pos 1 Tue | 0.590 | 1 | 0.057 | −0.281 | 0.213 | −0.262 | 0.580 | 0.315 | 0.496 | 0.672 |
| Pos 2 Thu | 0.012 | 0.057 | 1 | 0.185 | 0.278 | 0.348 | −0.030 | −0.231 | 0.040 | 0.274 |
| Pos 2 Fri | −0.009 | −0.281 | 0.185 | 1 | 0.269 | 0.055 | −0.087 | 0.236 | −0.367 | −0.006 |
| Pos 3 Sun | 0.408 | 0.213 | 0.278 | 0.269 | 1 | −0.182 | 0.454 | 0.174 | −0.212 | 0.474 |
| Pos 3 Sat | −0.470 | −0.262 | 0.348 | 0.055 | −0.182 | 1 | −0.396 | **−0.475** | −0.011 | −0.314 |
| Pos 4 Tue | 0.573 | 0.580 | −0.030 | −0.087 | 0.454 | −0.396 | 1 | 0.137 | 0.119 | 0.701 |
| Pos 4 Wed | 0.701 | 0.315 | −0.231 | 0.236 | 0.174 | −0.475 | 0.137 | 1 | 0.268 | 0.448 |
| Pos 5 Sat | 0.436 | 0.496 | 0.040 | −0.367 | −0.212 | −0.011 | 0.119 | 0.268 | 1 | 0.390 |
| Pos 5 Fri | **0.829** | 0.672 | 0.274 | −0.006 | 0.474 | −0.314 | 0.701 | 0.448 | 0.390 | 1 |

Bold values correspond to the maximum and minimum values. This is, the best correlation coefficient and the smallest correlation coefficient.

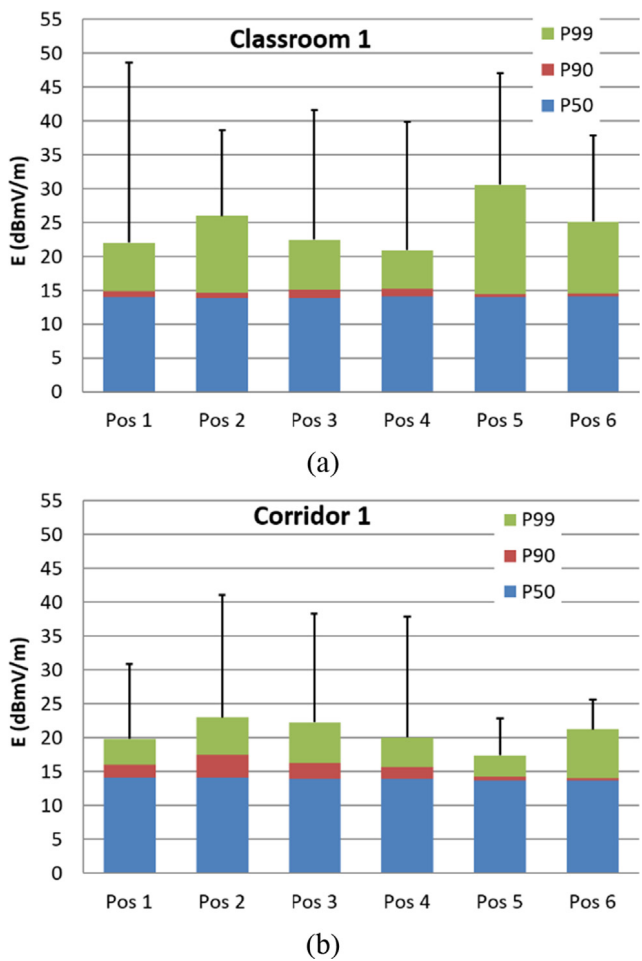

**Fig. 5.** Electric field levels E(dBmV/m) recorded with the spectrum analyzer at different positions (a) inside Classroom 1, (b) in Corridor 1.

### 3.2. 1-hour measurements

Inside a location (classroom or corridor), considerable variations in the measured field levels were detected due to the different placements of the measuring equipment. Classroom 1 was the classroom with the highest spatial variability of the measured signal, being differences in the maximum measured field with the spectrum analyzer up to 190.82 mV/m when placing the receiver at different positions. The P50, P90, P99 and maximum of the electric field values obtained in the different positions of this location are represented in Fig. 5a. Regarding the corridors, the highest variability due to the different positions was found in Corridor 1, where the maximum signal level varied up to 99.29 mV/m, the

P99 up to 6.69 mV/m and the P90 up to 2.46 mV/m when using the spectrum analyzer. Fig. 5b shows different statistics of the WiFi signal recorded in the different positions of this corridor.

Finally, the exposure assessment in each corridor (Cor) and classroom (Cla) when using the spectrum analyzer and the personal exposure meter are summarized in Table 7 and Table 8, respectively. Different percentiles and the mean electric field value were calculated taking into account the six positions inside a classroom or corridor. The maximum and the minimum WiFi sample level acquired in each location are also provided in the tables. Differences in the obtained results can be seen due to the different instrumentation used. Focusing on the data recorded by means of the spectrum analyzer, the minimum field strength value obtained in the six locations was below the detection limit of the exposure meter (5 mV/m). An overestimation could be produced by the EME Spy when there was low WiFi signal, since it cannot detect field levels lower than that limit. Furthermore, the personal exposure meter was recording a sample each 4 s while the spectrum analyzer recorded a sample per second, so more information about the signal could be obtained in this second case. This is important due to the signal transmission characteristics. WiFi signal is transmitted in the form of pulses of short duration, so a high signal variability in a short period of time is produced. Both equipment were not exactly in the same position, they were separated a minimum distance of 40 cm to fulfil the far field region requirement.

The maximum measured level by means of the spectrum analyzer was equal to 268.68 mV/m, and when using the EME Spy, a maximum level of 242 mV/m was recorded, both of them obtained in Classroom 1. Despite the differences in the results, both instruments are useful and suitable for assessing WiFi exposure variability at 2.4 GHz, as well as for comparing with exposure limits. Although the spectrum analyzer can provide more accurate results, personal exposure meters are useful to obtain information about the exposure distribution.

## 4. Comparison with wifi levels obtained in different measurement campaigns

Several measurement campaigns carried out by different authors were selected in order to compare the WiFi exposure levels obtained in different investigations. The instrumentation and the measurement scenarios in the different studies are described below. All the values were converted to electric field units (V/m) to allow the comparison with the exposure levels obtained in this work, as shown in Table 9.

Some studies in which data were recorded by means of personal exposure meters were selected. The EME Spy 120 was used in [31] and in [5]. In the first work, two different methods were used to calculate the statistics of the electric field when the measured level was below the detection limit of the instrument: the naïve

**Table 7**
Spectrum analyzer Results (mV/m).

|  | Cor 1 | Cor 2 | Cor 3 | Cla 1 | Cla 2 | Cla 3 |
|---|---|---|---|---|---|---|
| Min | 4.71 | 4.68 | 4.75 | 4.81 | 4.71 | 4.71 |
| Mean (SD) | 5.27 (0.23) | 4.99 (0.09) | 5.21 (0.08) | 5.36 (0.09) | 4.98 (0.08) | 5.19 (0.13) |
| P50$_{Av}$[a] (SD) | 4.97 (0.09) | 4.87 (0.02) | 4.99 (0.05) | 5.02 (0.04) | 4.86 (0.01) | 4.96 (0.08) |
| P90$_{Av}$[a] (SD) | 6.09 (0.84) | 5.25 (0.45) | 5.42 (0.18) | 5.51 (0.16) | 5.14 (0.18) | 5.85 (0.56) |
| P99$_{Av}$[a] (SD) | 10.98 (2.20) | 7.66 (1.33) | 11.06 (2.56) | 18.14 (7.65) | 7.76 (2.33) | 11.80 (7.88) |
| Max$_{Av}$[a] (SD) | 56.90 (36.54) | 24.15 (10.32) | 41.04 (14.16) | 146.01 (73.76) | 36.75 (19.94) | 74.93 (51.39) |
| Max | 113.15 | 46.98 | 69.29 | 268.68 | 72.78 | 183.98 |

SD: standard deviation.

[a] Mean values of the P50, P90, P99 and maximum levels of the electric field (mV/m) measured in the different positions of each place.

**Table 8**
Exposimeter Results (mV/m).

|  | Cor 1 | Cor 2 | Cor 3 | Cla 1 | Cla 2 | Cla 3 |
|---|---|---|---|---|---|---|
| Min | 5.00 | 17.00 | 5.00 | 13.00 | 8.00 | 5.00 |
| Mean (SD) | 15.62 (7.44) | 50.16 (12.96) | 16.24 (10.35) | 70.90 (28.57) | 19.35 (3.80) | 12.39 (1.62) |
| P50$_{Av}$[a] (SD) | 15.33 (7.70) | 47.83 (15.43) | 15.17 (11.13) | 67.33 (28.98) | 18.17 (3.80) | 11.67 (1.70) |
| P90$_{Av}$[a] (SD) | 22.32 (9.67) | 64.83 (11.45) | 24.52 (13.07) | 84.87 (31.83) | 25.33 (3.77) | 19.00 (3.45) |
| P99$_{Av}$[a] (SD) | 37.32 (13.96) | 72.84 (9.97) | 47.01 (22.95) | 115.03 (50.64) | 38.00 (2.58) | 37.00 (9.97) |
| Max$_{Av}$[a] (SD) | 89.33 (64.79) | 83.00 (3.83) | 84.83 (44.61) | 186.54 (31.66) | 46.17 (4.52) | 53.83 (26.70) |
| Max | 223.00 | 87.00 | 165.00 | 242.00 | 54.00 | 113.00 |

SD: standard deviation.

[a] Mean values of the P50, P90, P99 and maximum levels of the electric field (mV/m) measured in the different positions of each place.

**Table 9**
Electric field levels (V/m) obtained in different measurement campaigns.

| Ref | Mean | Median | Range | Description |
|---|---|---|---|---|
| [34] | 0.01–0.03 | – | – | Eme Spy 201 ExpoM-RF |
| [5] | 0.019–0.082 | – | – | Eme Spy 120/121 |
| [35] | 0.020 | – | 0.006–0.1 | Weighting factor |
| [31] ROS | 0.05 | 0.02 | NA–0.23 | Eme Spy 120 |
| [31] Naïve | 0.06 | 0.05 | 0.05–0.22 | Eme spy 120 |
| [20] | 0.060–0.114 | – | – | Radiation Meter |
| [8] | 0.077–0.118 | 0.000–0.013 | – | Max-hold |
| Our work | 0.005 | 0.005* | 0.004–0.408 | Analyzer 24 h |
| Our work | 0.005 | 0.005* | 0.005*–0.269 | Analyzer 1 h |
| Our work | 0.031 | 0.029 | 0.005–0.242 | Eme Spy 200 1 h |

* The calculated values were between 0.0045 and 0.0049 V/m, but when rounding to the nearest third decimal a value of 0.005 V/m is set.

approach, which was the one used in our work, and the robust regression on order statistics (ROS) method, which consists of calculating statistics by fitting an assumed distribution to the observed data when they are below the detection limit. In this way, data values below the detection limit could be obtained. Participants carried out an exposimeter during one week and the weekly statistics from 109 participants are given in Table 9. In [5] the electric field levels in various locations of five countries were assessed and the average field levels measured in offices are given in Table 9, since this is a similar environment to the one considered in our work. In [34] two different exposure meters were employed, the ExpoM-RF and the EME Spy 201, with a detection limit of 0.005 V/m at 2.4 GHz in order to measure electric fields in 5 different countries, selecting for the comparison the average exposure in university areas.

In [20] measurements in several schools were performed using a selective radiation meter and a tri-axial probe at 1.5 m above the ground. They performed some stationary measurements and also took samples while walking through the classrooms. For this comparison, the stationary measurements taken under the access point (average field value 0.114 V/m) and at the furthest student desk to the access point (average field level 0.060 V/m) were selected. The average values of the field strengths were recorded, but no information about the detector employed is given. A spectrum analyzer and a tri-axial probe were used in [35] to receive signal levels in 27 outdoor locations and 3 indoor places and the WiFi exposure ranged between 0.006 and 0.1 V/m. Results were obtained measuring the maximum field strength and weighting it by means of a weighting factor [23]. In [8] measurement campaigns were carried out in several bedrooms during nights, repeating the measurements during three different years. The instrumentation employed consisted of a spectrum analyzer configured for obtaining maximum power levels (max-hold) and two biconical antennas. The median values of the electric fields were 0.00 V/m in 2006, 0.006 V/m in 2009 and 0.013 V/m in 2012, while the average values were 0.077 V/m in 2006, 0.118 V/m in 2009 and

0.105 V/m in 2012. Significant differences were observed between the median and mean values, explained because of the high standard deviation in the mean results.

As shown in Table 9, the results obtained with the EME Spy 120/121 are higher than those acquired in our work or with the EME Spy 200. One reason is that the detection limit of the EME Spy 120/121 is 0.05 V/m, while the EME Spy 200 and 201 can detect field levels higher than 0.005 V/m, so the probability of overestimation in the first instrument is higher. When using spectrum analyzers, differences due to the equipment configurations can occur. We used the clear/write trace and the RMS detector to avoid overestimations of the signals, but for example in [8] max-hold measurements were taken. This is usually done when verifying compliance with the regulations in order to check the worst case scenario in terms of exposure levels. However, an increase in the average results is produced.

Finally, a comparison with the exposure limits given in the standards was done. The maximum sample level was obtained with the spectrum analyzer in Lab 2 and it was equal to 0.408 V/m. The exposure limit established in [36] for the general public in the 2.4 GHz frequency band is equal to 61 V/m, so the maximum measured sample is far below the reference limits.

## 5. Conclusions

This paper presents an evaluation of the WiFi signal levels in the 2.4 GHz frequency band inside the university. The first objective of this work was to perform measurements in real environments using the configuration of the spectrum analyzer developed by authors of the present paper and used previously in controlled conditions, where a good agreement between the data traffic and the measured levels was obtained [25]. The usefulness of this configuration lies in the acquisition of accurate and realistic results, performing only frequency domain measurements, since in order to increase the accuracy provided by personal exposure meters and max-hold measurements with spectrum analyzers, some authors introduced weighting factors, but in these cases additional measurements are needed to calculate the weighting factors.

A measurement campaign based on recordings of 24 h and 1 h durations was carried out using a spectrum analyzer appropriately configured (see Table 2) together with a tri-axial antenna system, in order to assess the temporal and spatial variability of WiFi exposure levels. A personal exposure meter was also employed for the 1-hour measurements and the differences in the results obtained with both instruments were explained.

Finally, exposure levels due to WiFi signals measured in other studies were compared, detailing the equipment and environments of these measurement campaigns, concluding that despite the differences in the results due to the use of different equipment or configurations, all the values presented were far below the exposure limits. The methods shown in the different measurement campaigns are useful to obtain the field distribution and to check compliance with exposure limits. However, if more accurate and robust results are required, professional equipment appropriately configured should be use.

Regarding the statistical analysis, it was observed that as signal levels can vary significantly, many times the mean values do not provide enough information because of the high standard deviation of the results. Although the mean, median and maximum values are suitable for epidemiological studies and for checking compliance with exposure limits, when more information about the signal is desired, percentiles are more appropriate statistics, since they give information on exposure levels for the different percentages of the measurement time.

## Declaration of Competing Interest

The authors declared that there is no conflict of interest.

## Acknowledgment

This work has been financially supported by the Basque Government (IT1234-19), by the University of the Basque Country UPV/EHU under Grant Dokberri 2018-II (DOCREC18/36), and by the Spanish Government under the grant RTI2018-099162-B-I00 (MCIU/AEI/FEDER, UE).